\documentclass[prd,twocolumn,twoside,preprintnumbers,superscriptaddress,nofootinbib]{revtex4}
\usepackage{amsmath,slashed}
\usepackage{graphicx,graphics,color}
\usepackage{dcolumn}
\usepackage[hyperfootnotes=false]{hyperref}
\usepackage{xspace}
\usepackage{enumerate}
\usepackage{varwidth}

\newcommand{\Order}{\mathcal{O}}

\newcommand{\GeV}{\,\text{GeV}}
\newcommand{\TeV}{\,\text{TeV}}

\newcommand{\beq}{\begin{equation}}
\newcommand{\eeq}{\end{equation}}
\renewcommand{\Re}{\text{Re}\,}
\renewcommand{\Im}{\text{Im}\,}
\newcommand{\Br}{\text{Br}}
\newcommand{\Li}{\text{Li}}

\newcommand{\MU}{M_\Upsilon}

\allowdisplaybreaks[1]

\begin{document}

\preprint{PSI-PR-21-27, ZU-TH 56/21}
\title{Towards testing the magnetic moment of the tau at one part per million}

\author{Andreas Crivellin}
\affiliation{Physik-Institut, Universit\"at Z\"urich, Winterthurerstrasse 190, 8057 Z\"urich, Switzerland}
\affiliation{Paul Scherrer Institut, 5232 Villigen PSI, Switzerland}
\author{Martin Hoferichter}
\affiliation{Albert Einstein Center for Fundamental Physics, Institute for Theoretical Physics, University of Bern, Sidlerstrasse 5, 3012 Bern, Switzerland}
\author{J.~Michael Roney}
\affiliation{University of Victoria, Victoria, British Columbia, V8W 3P6, Canada}
\affiliation{Institute of Particle Physics (Canada)}

\begin{abstract}
 If physics beyond the Standard Model (BSM) explains the $4.2\sigma$ difference between the Standard-Model and measured muon anomalous magnetic moment, $a_{\mu}$, minimal flavor violation predicts a shift in the analog quantity for the $\tau$ lepton, $a_{\tau}$, at the $10^{-6}$ level, and even larger effects are possible in generic BSM scenarios such as leptoquarks. We show that this produces equivalent BSM deviations in the Pauli form factor, $F_2(s)$, at $s=(10\GeV)^2$, and report the  first complete two-loop prediction of $\Re F_2^\text{eff}(100\GeV^2)=-268.77(50)\times 10^{-6}$ for resonant $\tau$-pair production  in $e^+e^-\rightarrow \Upsilon(nS) \rightarrow$ $\tau^+\tau^-$, $n=1,2,3$.  $\Re F_2^\text{eff}$ can be measured from $e^-$-helicity-dependent transverse and longitudinal asymmetries in $\tau$-pair events, which requires a longitudinally polarized $e^-$ beam. 
 We discuss how Belle~II asymmetry measurements could probe $a_{\tau}^\text{BSM}$ at $10^{-6}$, assuming such a polarization upgrade of the SuperKEKB $e^+e^-$ collider, and conclude by outlining the next steps to be taken in theory and experiment along this new avenue for exploring realistic BSM effects in $a_\tau$.
\end{abstract}

\maketitle

\section{Introduction}

Searching for physics beyond the Standard Model (BSM)  in lepton anomalous magnetic moments, $a_\ell$, $\ell=e,\mu,\tau$, has a long tradition that dates back to Schwinger's famous prediction $a_\ell=(g-2)_\ell/2=\alpha/(2\pi)\simeq 1.16\times 10^{-3}$~\cite{Schwinger:1948iu} and its subsequent confirmation in experiment~\cite{Kusch:1948mvb}. For electrons and muons such precision tests have reached a level below $10^{-12}$ and $10^{-9}$, respectively. For the former, the comparison of the direct measurement~\cite{Hanneke:2008tm} and the SM prediction yields
\begin{align}
 a_e^\text{exp}-a_e^\text{SM}[\text{Cs}]&=-0.88(28)(23)[36]\times 10^{-12},\notag\\
 a_e^\text{exp}-a_e^\text{SM}[\text{Rb}]&=+0.48(28)(9)[30]\times 10^{-12},
\end{align}
depending on whether the fine-structure constant $\alpha$ is taken from Cs~\cite{Parker:2018vye} or Rb~\cite{Morel:2020dww} atom interferometry (the errors refer to $a_e^\text{exp}$, $\alpha$, and total, respectively). The $5.4\sigma$ tension between these measurements of $\alpha$ currently constitutes the biggest uncertainty, but, once resolved, further improvements in $a_e^\text{exp}$~\cite{Gabrielse:2019cgf} would allow one to probe $a_e$ at the level of $10^{-13}$ and beyond. 
On the theory side, 4-loop QED contributions are known semi-analytically~\cite{Laporta:2017okg}, while the $4.8\sigma$ tension between the numerical evaluations of the $5$-loop coefficient from Refs.~\cite{Aoyama:2019ryr,Volkov:2019phy} amounts to $6\times 10^{-14}$. However, each evaluation quotes an accuracy of $10^{-14}$, which is also the level at which hadronic uncertainties enter~\cite{Keshavarzi:2019abf} and thus defines the precision one may ultimately hope to reach for $a_e$.

For the muon, the experimental average~\cite{Bennett:2006fi,Abi:2021gix,Albahri:2021ixb,Albahri:2021kmg,Albahri:2021mtf} differs from the SM prediction~\cite{Aoyama:2020ynm} (mainly based on Refs.~\cite{Aoyama:2012wk,Aoyama:2019ryr,Czarnecki:2002nt,Gnendiger:2013pva,Davier:2017zfy,Keshavarzi:2018mgv,Colangelo:2018mtw,Hoferichter:2019gzf,Davier:2019can,Keshavarzi:2019abf,Hoid:2020xjs,Kurz:2014wya,Melnikov:2003xd,Colangelo:2014dfa,Colangelo:2014pva,Colangelo:2015ama,Masjuan:2017tvw,Colangelo:2017qdm,Colangelo:2017fiz,Hoferichter:2018dmo,Hoferichter:2018kwz,Gerardin:2019vio,Bijnens:2019ghy,Colangelo:2019lpu,Colangelo:2019uex,Blum:2019ugy,Colangelo:2014qya})
by
\beq
\label{Deltaamu}
a_\mu^\text{exp}-a_\mu^\text{SM}=2.51(59)\times 10^{-9},
\eeq
with an uncertainty that derives to about equal parts from experiment and theory. Future runs of the Fermilab experiment are projected to reach a precision of $1.6\times 10^{-10}$~\cite{Grange:2015fou},  while progress on the theory side will require the emerging tension between $e^+e^-\to \text{hadrons}$ data, on which the consensus value from Ref.~\cite{Aoyama:2020ynm} is based, and the lattice-QCD calculation~\cite{Borsanyi:2020mff}, see Refs.~\cite{Lehner:2020crt,Crivellin:2020zul,Keshavarzi:2020bfy,Malaescu:2020zuc,Colangelo:2020lcg}, to be resolved, potentially including an independent measurement at the proposed MUonE experiment~\cite{MUonE:LoI,Banerjee:2020tdt}.
Even though further experimental improvements may be possible, e.g., with the proposed High Intensity Muon Beam at PSI~\cite{Aiba:2021bxe}, it seems hard to move beyond  a level of $10^{-10}$ in precision. 

Compared to these precision tests, information on $a_\tau$ is minimal, with the range
\beq
-0.052 < a_\tau^\text{exp} < 0.013 \quad [2\sigma],
\eeq
extracted from $e^+e^-\to e^+e^-\tau^+\tau^-$ at LEP2~\cite{DELPHI:2003nah}. In a global analysis of LEP and SLD data~\cite{L3:1998lhr,SLD:1999uov,ALEPH:2000fhd} in effective field theory (EFT), a tighter limit on BSM contributions can be derived~\cite{Gonzalez-Sprinberg:2000lzf}
\beq
-0.007 < a_\tau^\text{BSM} < 0.005 \quad [2\sigma],
\eeq
still well above the size of Schwinger's $1$-loop QED result. In contrast, scaling the tension~\eqref{Deltaamu} with $(m_\tau/m_\mu)^2$ would imply $ a_\tau^\text{BSM} \simeq  
0.7\times 10^{-6}$. Since this is of the order of the SM electroweak (EW) contribution, $a_\tau^\text{EW}\simeq 0.5\times 10^{-6}$~\cite{Eidelman:2007sb}, this also sets the scale at which BSM effects can reasonably arise, i.e., a few times~$10^{-6}$ as we will show below using a leptoquark (LQ) example.\footnote{The quadratic scaling assumes minimal flavor violation (MFV)~\cite{Chivukula:1987fw,Hall:1990ac,Buras:2000dm,DAmbrosio:2002vsn}. However, MFV is challenged by the recent flavor anomalies (see Ref.~\cite{Fischer:2021sqw} for an overview) and more generic models are well motivated~\cite{Crivellin:2021rbq}.} A precision at this level is thus required for a meaningful test of $a_\tau$. 

Unfortunately, reaching this level in $a_\tau$ is extremely challenging: alternative methods using radiative $\tau$ decays~\cite{Eidelman:2016aih}, channeling~\cite{Kim:1982ry} in a bent crystal~\cite{Samuel:1990su,Fomin:2018ybj,Fu:2019utm}, or $\gamma p$~\cite{Koksal:2017nmy,Gutierrez-Rodriguez:2019umw} and heavy-ion~\cite{Beresford:2019gww,Dyndal:2020yen,ATLAS:2022ryk,CMS:2022arf} reactions at the LHC have been proposed. However, the only projections that reach down to $10^{-6}$ (for the statistical error) have been obtained from $e^+e^-\to\tau^+\tau^-$ at the $\Upsilon$ resonances~\cite{Bernabeu:2007rr,Bernabeu:2008ii} (earlier, also the threshold region was considered~\cite{Brodsky:1995ds}). The key idea is that at the resonance other diagrams than those mediated by $s$-channel exchange, e.g., box diagrams, will be suppressed, leading to an enhanced sensitivity to the Pauli form factor $F_2$ at the $\Upsilon$ mass, which can then be converted to a constraint on $a_\tau$. Since a measurement of an absolute cross section at $10^{-6}$ would be extremely challenging~\cite{Chen:2018cxt,Kaiser:2021erl}, with the dominant uncertainties of systematic origin, we concentrate on the transverse and longitudinal asymmetries suggested in Ref.~\cite{Bernabeu:2007rr}, which become accessible if polarized beams are available. In addition, even if the experimental precision can be reached, to test $a_\tau$ at $10^{-6}$ we need a theoretical description valid at the same level, which, as we will discuss next, requires $2$-loop accuracy.

\begin{table}[t]
	\renewcommand{\arraystretch}{1.3}
	\centering
	\begin{tabular}{l r r}
		\toprule
		 & $s=0$ & $s=(10\GeV)^2$\\\colrule
$1$-loop QED & $1161.41$ & $-265.90+246.48i$\\
$e$ loop & $10.92$ & $-2.43+2.95i$\\
$\mu$ loop & $1.95$ & $-0.34+0.92i$\\
$\tau$ loop & $0.08$ & $0.06+0.07i$\\
$2$-loop QED (mass & $-1.77$ & IR divergent\\
 independent, incl.\ $\tau$ loop) & &\\
 sum QED &  $1172.51$ & IR divergent\\
 HVP & $3.33$ & $-0.33+1.93i$\\\colrule
 sum of the above & $1175.84$ &\\
 QED (incl.\ $3$-loop)~\cite{Eidelman:2016aih} & $1173.24(2)$ &\\
 HVP~\cite{Keshavarzi:2019abf} & $3.328(14)$ &\\
 EW~\cite{Eidelman:2016aih} & $0.474(5)$ &\\
 total~\cite{Keshavarzi:2019abf} & $1177.171(39)$ &\\
		\botrule
	\end{tabular}
	\caption{Contributions to $F_2(s)$ in units of $10^{-6}$. Examples for the topologies are shown in Fig.~\ref{fig:diagrams}.}  
	\label{tab:F2}
\end{table}
 
\section{Pauli form factor and \texorpdfstring{$\boldsymbol{(g-2)_\tau}$}{}}

We work with the Dirac and Pauli form factors $F_{1,2}(s)$ in the standard convention
\beq
\label{Fidef}
\langle p'|j^{\mu}|p\rangle
=e \bar u(p')\Big[\gamma^\mu F_1(s)+\frac{i\sigma^{\mu\nu}q_\nu}{2m_\tau} F_2(s)\Big]u(p),
\eeq
with $F_1(0)=1$, $F_2(0)=a_\tau$, and $q=p'-p$. 
To gauge the precision requirements we first consider the decomposition of $a_\tau^\text{SM}$ following Ref.~\cite{Eidelman:2016aih}, with a recent update for hadronic vacuum polarization (HVP)~\cite{Keshavarzi:2019abf}. The values of $F_2(0)$ given in Table~\ref{tab:F2} show that the infrared (IR) enhancement by $\log \frac{m_\ell}{m_\tau}$ increases the contribution from the electron loop to well above $10^{-6}$, and also the muon-loop, HVP, and even the mass-independent $2$-loop QED contributions are non-negligible (note that for the $\tau$ loop there is an accidental cancellation). Including the EW correction, the deficit to the full SM prediction becomes $0.86\times 10^{-6}$, of which $0.73\times 10^{-6}$ is from $3$-loop QED (dominated by electron light-by-light scattering) and the rest from higher-order hadronic effects. This decomposition shows that $2$-loop effects are indeed required to claim this precision, and that when $a_\tau$ is truncated at this level, the error stays below $10^{-6}$. 

Accordingly, the transition from $s=0$ to $s=\MU^2$ has to proceed at the same order. The imaginary parts of the QED form factors have been known since Refs.~\cite{Barbieri:1972as,Barbieri:1972hn},  and the real parts can be derived, e.g., from a dispersion relation, see Refs.~\cite{Mastrolia:2003yz,Bonciani:2003ai} for explicit expressions. The mass-dependent contributions  can be treated as described in App.~\ref{app:mass-dependent}, leading to the results shown in Table~\ref{tab:F2} for $s=(10\GeV)^2$. One sees that for the light degrees of freedom (electron, muon, and hadrons) the real part is suppressed significantly and even the sign changes when moving from zero momentum transfer to $s=(10\GeV)^2$, while the pure QED contribution becomes IR divergent. In contrast, the EW contribution is only modified by negligible corrections $\propto s/M_Z^2$ (see, e.g., Refs.~\cite{Papavassiliou:1989zd,
Bernabeu:2002nw} for the gauge-invariant definition in non-Abelian theories), and likewise a potential heavy BSM contribution to $a_\tau$ would remain unaffected up to tiny dimension-$8$ effects.  

\begin{figure}[t!]
	\centering
	\includegraphics[width=\linewidth]{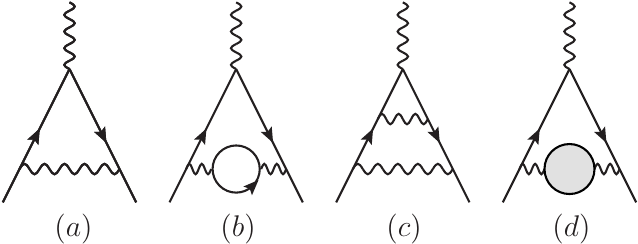}
	\caption{Representative diagrams contributing to the QED form factors: (a) $1$-loop QED, (b) lepton loops, (c) $2$-loop QED, (d) HVP; the gray blob denotes the hadronic 2-point function.}
	\label{fig:diagrams}
\end{figure}

The presence of the IR divergence signals that away from $s=0$ the form factor $F_2$ alone does not describe a physical process, and only the combination with soft radiation becomes observable. Following Ref.~\cite{Mastrolia:2003yz} we will introduce as regulator a finite photon mass $\lambda$ (see Ref.~\cite{Bonciani:2003ai} for the results in dimensional regularization) and turn next to $e^+e^-\to\tau^+\tau^-$.

\section{\texorpdfstring{$\boldsymbol{e^+e^-\to\tau^+\tau^-}$}{} cross section and asymmetries}

The differential cross section for $e^+e^-\to \tau^+\tau^-$ takes the form
\begin{align}
\label{cross_section}
\frac{d\sigma}{d\Omega}&=\frac{\alpha^2\beta}{4s}\bigg[\big(2-\beta^2\sin^2\theta\big)\Big(|F_1|^2-\gamma^2|F_2|^2\Big)\notag\\
&+4\Re\big(F_1F_2^*\big)+2(1+\gamma^2)|F_2|^2\bigg],
\end{align}
with center-of-mass scattering angle $\theta$, $\beta=\sqrt{1-4m_\tau^2/s}$, and $\gamma=\sqrt{s}/(2m_\tau)$, when assuming the general form-factor decomposition~\eqref{Fidef}. Extracting constraints on $a_\tau^\text{BSM}$ from $e^+e^-\to\tau^+\tau^-$ thus proceeds via the term $\Re(F_1 F_2^*)$, which is the only one sensitive to $2$-loop effects in $F_2$, while corrections from $|F_2|^2$ do need to be kept. Disentangling $\Re(F_1 F_2^*)$ in Eq.~\eqref{cross_section} from the dominant $|F_1|^2$ term by means of the angular dependence is possible if semileptonic decays of the $\tau$ are considered~\cite{Bernabeu:2007rr,Tsai:1971vv,Kuhn:1993ra}, which allows one to reconstruct the $\tau$ production plane and direction of flight, and thereby the scattering angle. However, a determination of $F_2$ at the level of $10^{-6}$ not only requires an absolute cross-section measurement at that accuracy, but also full radiative corrections at $2$-loop order, including double bremsstrahlung to remove the IR divergences in $F_1$. 

Instead, we therefore turn to the transverse and longitudinal asymmetries $A_T^\pm$, $A_L^\pm$ constructed in Ref.~\cite{Bernabeu:2007rr}, with the key idea being to use polarization to disentangle the different contributions to the cross section. Retaining the information on the spins $s_\pm$ of the $\tau^\pm$ and the electron helicity $\lambda$, one has
\begin{align}
 \frac{d\sigma^{S\lambda}}{d\Omega}&=\frac{\alpha^2\beta}{16s}\big\{s_y Y+\lambda\Big[s_x X+s_zZ\Big]\Big\},\quad\! s_i\equiv (s_++s_-)_i,\notag\\
 X&=\frac{\sin\theta}{\gamma}\Big[|F_1|^2+(1+\gamma^2)\Re(F_2 F_1^*)+\gamma^2|F_2|^2\Big],\notag\\
 Y&=\frac{\gamma\beta^2}{2}\sin(2\theta)\Im(F_2 F_1^*),\quad
 Z=\cos\theta |F_1+F_2|^2,
\end{align}
while the spin-independent term (summed over $s_\pm$, $\lambda$) reproduces Eq.~\eqref{cross_section} (we will denote the cross section summed over $\lambda$ by $d\sigma^S$). Accordingly, without polarization only $\Im F_2$ is accessible from the spin-dependent terms~\cite{Bernabeu:2004ww,Bernabeu:2006wf}, which can be measured if the $\tau^\pm$ are reconstructed from semileptonic decays. To this end, these decays are characterized via~\cite{Bernabeu:2007rr}
\beq
\mathbf{n}_\pm^*=\mp \alpha_\pm\big(\sin\theta_\pm^*\cos\phi_\pm,\sin\theta_\pm^*\sin\phi_\pm,\cos\theta_\pm^*),
\eeq
where $\phi_\pm$ and $\theta_\pm^*$ are the azimuthal and polar angles of the produced hadron $h^\pm$ (e.g., $h=\pi,\rho$) in the $\tau^\pm$ rest frame and $\alpha_\pm \equiv \frac{m^2_{\tau}-2m^2_{h^{\pm}}}{m^2_{\tau}+2m^2_{h^{\pm}}}$ is the polarization analyzer~\cite{Bernabeu:1993er}. 
Reference~\cite{Bernabeu:2007rr} then suggests to measure the asymmetries
\beq
A_T^\pm=\frac{\sigma_R^\pm-\sigma_L^\pm}{\sigma},\qquad 
A_L^\pm=\frac{\sigma_{\text{FB},\,R}^\pm-\sigma_{\text{FB},\,L}^\pm}{\sigma},
\eeq
where $\sigma$ denotes the total cross section (for semileptonic $\tau$ decays in the final state), and the transverse and longitudinal differences are constructed as follows: first, we define the helicity difference as
\beq
d\sigma^{S}_\text{pol}=\frac{1}{2}\Big(d\sigma^{S\lambda}\big|_{\lambda=1}
-d\sigma^{S\lambda}\big|_{\lambda=-1}\Big).
\eeq
For $A_T^\pm$ we then integrate over $d\Omega$ and all angles except for $\phi_\pm$:
\beq
\sigma_R^\pm = \int_{-\pi/2}^{\pi/2}d\phi_\pm \frac{d\sigma^S_\text{pol}}{d\phi_\pm},\quad
\sigma_L^\pm = \int_{\pi/2}^{3\pi/2}d\phi_\pm \frac{d\sigma^S_\text{pol}}{d\phi_\pm}.
\eeq
For $A_L^\pm$ we instead use a forward--backward (FB) integration over $dz=d\!\cos\theta$,
\beq
\sigma_\text{FB}^S=\int_0^1dz\frac{d\sigma^S}{dz}-\int_{-1}^0dz\frac{d\sigma^S}{dz},
\eeq
and then integrate over all angles except for $\theta_\pm^*$:
\beq
\sigma_{\text{FB},\,R}^\pm = \int_{0}^{1}dz_\pm^* \frac{d\sigma^S_\text{FB,\,pol}}{dz_\pm^*},\quad \sigma_{\text{FB},\,L}^\pm = \int_{-1}^{0}dz_\pm^* \frac{d\sigma^S_\text{FB,\,pol}}{dz_\pm^*}.
\eeq
In the difference 
\beq
\label{ATL}
A_T^\pm -\frac{\pi}{2\gamma}A_L^\pm=\mp\alpha_\pm \frac{\pi^2\alpha^2\beta^3\gamma}{4s\sigma}\big[\Re(F_2 F_1^*)+|F_2|^2\big]
\eeq
the dominant contribution from $|F_1|^2$ cancels, isolating the desired effect from $F_2$. We reproduce the result from Ref.~\cite{Bernabeu:2007rr} by truncating Eq.~\eqref{ATL} at $1$-loop order, i.e., $\Re(F_2 F_1^*)\to \Re F_2$, $|F_2|^2\to 0$, and $\sigma\to\sigma_0=2\pi\alpha^2\beta(3-\beta^2)/(3s)$. To reach a level of $10^{-6}$, however, we need to evaluate Eq.~\eqref{ATL} at $2$-loop order, including radiative corrections both in the numerator and denominator. 
In fact, we can define an effective $\Re F_2^\text{eff}$ as
\beq
\label{RF2eff}
\Re F_2^\text{eff}=\mp \frac{8(3-\beta^2)}{3\pi\gamma\beta^2\alpha_\pm}\Big(A_T^\pm -\frac{\pi}{2\gamma}A_L^\pm\Big),
\eeq
which can be determined in experiment, and work out the corrections to relate this quantity to constraints on $a_\tau^\text{BSM}$ in the following.

\begin{table}[t]
	\renewcommand{\arraystretch}{1.3}
	\centering
	\begin{tabular}{l r r}
		\toprule
		 & $s=0$ & $s=(10\GeV)^2$\\\colrule
		 $1$-loop QED & $1161.41$ & $-265.90$\\
		 $e$ loop & $10.92$ & $-2.43$\\
		 $\mu$ loop & $1.95$ & $-0.34$\\
		 $2$-loop QED (mass ind.) & $-0.42$ & $-0.24$\\
		 HVP & $3.33$ & $-0.33$\\
		 EW & $0.47$ & $0.47$\\\colrule
		 total & $1177.66(86)$ & $-268.77(50)$\\ 
		\botrule
	\end{tabular}
	\caption{Contributions to $\Re F_2^\text{eff}(s)$, see Eq.~\eqref{Reff_exp_fin}, in units of $10^{-6}$. The values for $s=0$ are given for illustration only, by formal evaluation of Eq.~\eqref{Reff_exp_fin}, but since $\Re F_2^\text{eff}(s)$ is defined in terms of cross sections, values below $s=4m_\tau^2$ are not physical. The uncertainty at $s=0$ quantifies the size of $3$-loop effects in $a_\tau$, the one at $(10\GeV)^2$ is scaled by the suppression seen for the mass-independent $2$-loop QED contribution.}  
	\label{tab:F2eff}
\end{table}

\section{Radiative corrections}

Writing $F_i=F_i^{(0)}+\frac{\alpha}{\pi}F_i^{(1)}+\big(\frac{\alpha}{\pi}\big)^2F_i^{(2)}+\Order(\alpha^3)$, we have
\begin{align}
\label{Reff_exp}
 \Re F_2^\text{eff}&=\frac{\alpha}{\pi}\Re F_2^{(1)}+ \Big(\frac{\alpha}{\pi}\Big)^2\Big[\Re F_2^{(2)}
 +\Re F_2^{(1)}\Re F_1^{(1)}\notag\\
 &+\Im F_2^{(1)}\Im F_1^{(1)}+|F_2^{(1)}|^2\Big]\notag\\
 &+\Big(\frac{\sigma^0}{\sigma}-1\Big)\frac{\alpha}{\pi}\Re F_2^{(1)} + \Order\big(\alpha^3\big),
\end{align}
with
\beq
\label{sig0sig}
\frac{\sigma^0}{\sigma}-1=\frac{\alpha}{\pi}\Big[-2\Re F_1^{(1)}-\frac{6}{3-\beta^2}\Re F_2^{(1)}\Big]+\Order\big(\alpha^2\big),
\eeq
and the effect of a potential BSM contribution $a_\tau^\text{BSM}$ would first manifest itself as a modification of $\Re F_2^{(2)}$. IR divergences occur in $\Re F_2^{(2)}$, $\Re F_1^{(1)}$, and $\Im F_1^{(1)}$, which thus requires the inclusion of bremsstrahlung diagrams, see App.~\ref{sec:bremsstrahlung}. However, a similar IR divergence also occurs in Eq.~\eqref{sig0sig}, to the effect that in the end the corrections cancel in $\Re F_2^\text{eff}$, which can therefore be written in the form
\begin{align}
\label{Reff_exp_fin}
 \Re F_2^\text{eff}&=\frac{\alpha}{\pi}\Re F_2^{(1)}+ \Big(\frac{\alpha}{\pi}\Big)^2\Big[\Re F_2^{(2)}
 -\Re F_2^{(1)}\Re F_1^{(1)}\notag\\
 &+\Im F_2^{(1)}\big(\Im F_1^{(1)}+\Im F_2^{(1)}\big)\notag\\
 &-\frac{3+\beta^2}{3-\beta^2}\big(\Re F_2^{(1)}\big)^2\Big]+ \Order\big(\alpha^3\big).
\end{align}
The numerical contributions are listed in Table~\ref{tab:F2eff}. First, we show the comparison at $s=0$, in which case $\Re F_2^\text{eff}(0)=a_\tau+\frac{1}{4}\big(\frac{\alpha}{\pi}\big)^2$, reducing the mass-independent $2$-loop QED contribution to below $10^{-6}$. At $s=(10\GeV)^2$, we find that all corrections beyond the electron loop are already suppressed to this level as well, suggesting that the resulting SM prediction should be quite robust.\footnote{For the $2$-loop mass-independent QED contribution, there is a significant cancellation among the finite parts of the (separately IR divergent) terms involving $\Re F_2^{(2)}$, $\Re F_1^{(1)}$, and $\Im F_1^{(1)}$.} If the BSM scale is large compared to $s$, the comparison to the measured value of $\Re F_2^\text{eff}$ directly provides the constraint on $a_\tau^\text{BSM}$, while in the case of light new degrees of freedom, the EFT treatment no longer applies and the constraints would become model dependent. 

\section{Towards ppm precision}

The previous discussion assumes that corrections beyond the direct $s$-channel diagrams are negligible at the resonance, where the sensitivity to $F_2$ is enhanced by~\cite{Bernabeu:2007rr}
\beq
|H(M_\Upsilon)|^2=\Big(\frac{3}{\alpha}\Br(\Upsilon\to e^+e^-)\Big)^2\simeq 100,
\eeq
for $\Upsilon(nS)$, $n=1,2,3$. Unfortunately, the same idea does not work on the $\Upsilon(4S)$ resonance, where most of the running of a potential SuperKEKB upgrade with polarized electrons~\cite{Roney:2019til} would be anticipated, since there $\tau^+\tau^-$ pairs are not produced resonantly.
In addition, in practice the unavoidable spread in beam energies counteracts the resonance enhancement to the extent that for a typical spread, continuum $\tau^+\tau^-$ pairs outnumber resonant ones by almost a factor 10 at the $\Upsilon(3S)$ resonance~\cite{BaBar:2020nlq}.
In a similar vein, the precision of the subtraction in Eq.~\eqref{RF2eff} is limited by the uncertainty in $\gamma$,  arising from the uncertainties in the
mass of the $\Upsilon(1S)$, which is used to calibrate the center-of-mass energy in the machine, and $m_\tau$~\cite{ParticleDataGroup:2020ssz},
currently allowing for a precision of $1\times 10^{-5}$. Our study therefore motivates improved measurements of those quantities.

Assuming $40\,\text{ab}^{-1}$ of $e^+e^- \to \tau^+\tau^-$ data, with $60\%$ selection efficiency of the semileptonically decaying $\tau^\pm$, the statistical error on $\Re F_2^\text{eff}$ would become 
$1\times 10^{-5}$, while the 
dominant detector systematic uncertainties cancel in the asymmetries $A_{T,L}^\pm$. 
The path towards eventually constraining $a_\tau^\text{BSM}$ at the $10^{-6}$ level will thus require higher precision measurements of $m_{\tau}$ and  $M_{\Upsilon(1S)}$ as well as higher statistics. The latter could be obtained by including nonresonant data, although
the interpretation of such a measurement requires significant investments in theory development, to derive a SM prediction in analogy to Table~\ref{tab:F2eff}. In particular, the consideration of box diagrams~\cite{Kaiser:2021erl}, likely even at $2$-loop order, would become imperative, but this does not appear unrealistic given that similar corrections are currently being worked out in the context of MUonE~\cite{Banerjee:2020tdt}. 

\section{BSM scenarios}

Among models with MFV, the minimal supersymmetric SM (see, e.g., Ref.~\cite{Stockinger:2006zn} for a review on $g-2$), is still very popular, despite missing direct signals for supersymmetric partners~\cite{Lari:2020vke}. In such cases, $a_\tau^\text{MFV}=m_\tau^2/m_\mu^2 a_\mu^\text{MFV}\simeq 280\, a_\mu^\text{MFV}$ is predicted. The current discrepancy in $a_\mu$, Eq.~\eqref{Deltaamu}, then translates into 
\begin{equation}
\label{atauMFV}
    a_\tau^\text{MFV}=7.1(1.7)\times 10^{-7},
\end{equation}
indeed at the level of $10^{-6}$, which could be within reach of a SuperKEKB polarization upgrade.

Furthermore, it is well possible that MFV is not realized in nature and, in fact, the MFV paradigm is challenged by the anomalies in semi-leptonic $B$ decays, which rather point towards less-minimal flavor violation~\cite{Calibbi:2015kma,Barbieri:2016las,Bordone:2017bld,Fuentes-Martin:2019mun,Calibbi:2019lvs,Dorsner:2019itg,Fajfer:2021cxa,Marzocca:2021miv} or even an anarchic flavor structure (see, e.g., Refs.~\cite{Crivellin:2017zlb,Crivellin:2018yvo}). In particular, the couplings to $\tau$ leptons are much larger than to light leptons if one aims at explaining $R(D^{(*)})$. In this context, LQs (see Ref.~\cite{Dorsner:2016wpm} for a review) are particularly interesting as they can explain the hints for lepton flavor universality violation~\cite{Crivellin:2021sff}. Among the ten possible representations under the SM gauge group~\cite{Buchmuller:1986zs}, two can account for $a_\mu$ via a top-quark-mass chirally enhanced effect~\cite{Djouadi:1989md,Chakraverty:2001yg,Cheung:2001ip,ColuccioLeskow:2016dox} and they thus also have the potential to give rise to sizable contributions to $a_\tau$, beyond the MFV expectation~\eqref{atauMFV}. Interestingly, both of these representations also affect $R(D^{(*)})$~\cite{Bauer:2015knc,Becirevic:2016yqi}, motivating sizable couplings to $\tau$ leptons. Further, the $SU(2)_L$ singlet $S_1$ has the advantage of providing an explanation~\cite{Carvunis:2021dss} of the tension in $\Delta A_\text{FB}$~\cite{Bobeth:2021lya} in $B\to D^*\ell\nu$~\cite{Belle:2017rcc,Belle:2015pkj}.

\nocite{Arnan:2019olv,Crivellin:2020tsz,Crivellin:2020mjs}

\begin{figure}[t!]
	\centering
	\includegraphics[width=\linewidth]{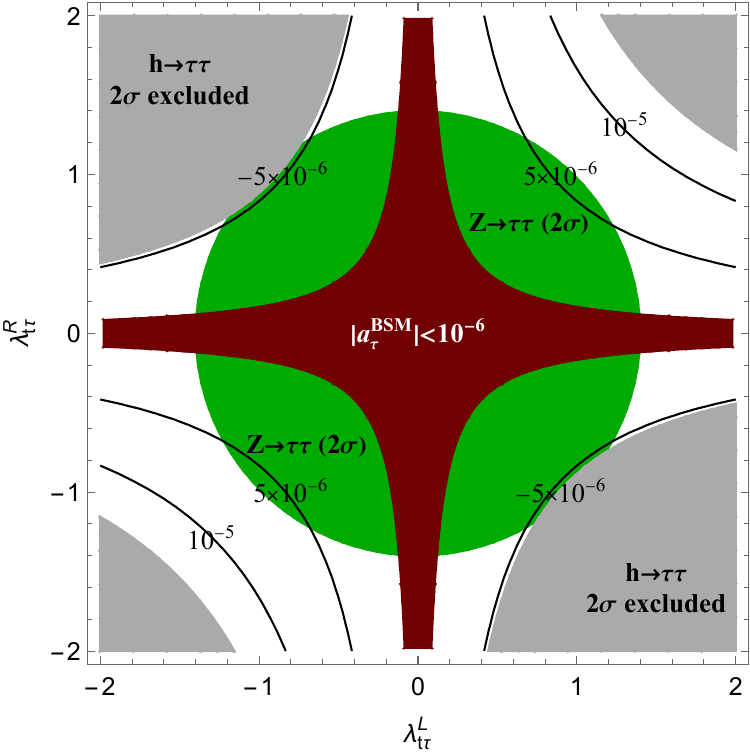}
	\caption{Illustration of the discovery potential of $a_\tau$ in the $S_1$ LQ model, taking into account the constraints from $h\to \tau^+\tau^-$ (gray, excluded) and $Z\to \tau^+\tau^-$ (green, allowed), with contour lines indicating the respective value of $a_\tau^\text{BSM}$.  The asymmetry in the $h\to\tau\tau$ constructive and destructive exclusion regions  originates from the current $1\sigma$ upward fluctuation in the data. The LQ mass is set to $M=2\TeV$, which is compatible with the bounds from direct searches~\cite{ATLAS:2019qpq,CMS:2018oaj}.}
	\label{fig:atauLQ}
\end{figure}

Therefore, we illustrate the potential for BSM effects in $a_\tau$ with $S_1$ as an example. As in all models with chiral enhancement~\cite{Crivellin:2021rbq}, $h\to \tau^+\tau^-$ and $Z\to \tau^+\tau^-$ are affected~\cite{ColuccioLeskow:2016dox,Arnan:2019olv,Crivellin:2020tsz}.\footnote{Also $Z\to \nu_\tau\bar \nu_\tau$ and $W\to\tau\bar\nu_\tau$ couplings receive, in general, loop contributions. However, for $S_1$ they prove to be subleading~\cite{Crivellin:2020mjs}.} Denoting the LQ mass by $M$ and the relevant couplings by $\lambda_{t\tau}^{R/L}$, see App.~\ref{sec:S1}, we find the constraints shown in Fig.~\ref{fig:atauLQ}, using the bounds on the branching fraction for $h\to\tau^+\tau^-$, $\Br\left[h \to \tau ^ +\tau ^ - \right]/\Br\left[ h \to \tau ^ + \tau ^ - \right]|_\text{SM}=1.15_{ - 0.15}^{ + 0.16}$ from the LHC~\cite{ParticleDataGroup:2020ssz,ATLAS:2018ynr,CMS:2018vqh,ATLAS:2016neq} and on the axial-vector coupling $g_A^\tau/g_A^\tau|_{\text{SM}}=1.00154(128)$ from LEP~\cite{ALEPH:2005ab}, respectively.
As one can see, a sizable effect in $a_\tau$, above the $10^{-6}$ level, is excluded neither by $h\to \tau^+\tau^-$ nor by $Z\to \tau^+\tau^-$, highlighting the discovery potential of SuperKEKB with polarization upgrade. At the same time, Belle II with polarized beams could also improve on the measurement of the (off-shell) $Z$--$\tau$--$\tau$ coupling, thus leading to interesting synergies in a future polarization program.

\section{Conclusions and outlook}

In this paper we studied a realistic and promising path towards testing the magnetic moment of the $\tau$ at one part per million. This level of precision is necessary to derive meaningful constraints on BSM physics, a conclusion that results 
both from an MFV-like scaling of the tension in $a_\mu$ and concrete models beyond MFV. As an example for the latter, we showed that leptoquark models can give rise to effects of the order of several times $10^{-6}$ without violating bounds from other precision observables. 

First, we argued that the decomposition of the SM prediction for $a_\tau$ implies that extractions from $e^+e^-\to \tau^+\tau^-$ need to proceed at $2$-loop accuracy and worked out the SM prediction for the case in which data are taken on the $\Upsilon(nS)$, $n=1,2,3$, resonances, where $\tau^+\tau^-$ pairs can be produced resonantly via their decays. Experimentally, the required precision will be difficult to achieve with absolute cross-section measurements, rendering the asymmetries suggested in Ref.~\cite{Bernabeu:2007rr} the most promising observables, whose measurement could be realized  at a SuperKEKB upgrade with polarized electrons. 

Ultimately, the path towards $10^{-6}$ 
requires improved measurements of $m_{\tau}$ and  $M_{\Upsilon(1S)}$, as well as  
collecting high-statistics samples that are dominated by
nonresonant $\tau$ pairs. 
 In this case, substantial investment in theory would be required to match the experimental efforts, but this would be well justified: given the hints for BSM contributions in $a_\mu$, such a program  would provide by far the most realistic avenue towards commensurate precision tests in $a_\tau$, further strengthening the physics case for a SuperKEKB polarization upgrade.

\begin{acknowledgments}
Support by the Swiss National Science Foundation, under Project No.\ PP00P21\_76884 (A.C.) and PCEFP2\_181117 (M.H.), is gratefully acknowledged, as is support from the Natural Sciences and Engineering Research Council of Canada (J.M.R.). 
\end{acknowledgments}

\appendix

\section{Mass-dependent \texorpdfstring{$\boldsymbol{2}$-loop}{-loop} contributions}
\label{app:mass-dependent}

The contribution from the lepton loops to $F_2(s)$ can be written in the form
\begin{align}
\label{F2lepton}
 F_2^\ell(s)&=-\frac{\alpha}{\pi}\int_0^1 dx\, (1-x)\int_0^{1-x}dy \frac{m_\tau^2}{s_{xy}}\bar\Pi_\ell(s_{xy}),\notag\\ 
 \bar \Pi_\ell(s)&=\frac{2\alpha}{\pi}\int_0^1dx\,x(1-x)\log\Big[1-x(1-x)\frac{s}{m_\ell^2}\Big],\notag\\
 s_{xy}&=-\frac{(1-x)^2}{x}m_\tau^2+\frac{y(1-x-y)}{x}s,
\end{align}
which in the limit $s\to 0$ indeed reduces to the expected expression
\beq
\label{F2lepton0}
F_2^\ell(0)=\frac{\alpha}{\pi}\int_0^1 dx\, x\,\bar\Pi_\ell\bigg(-\frac{(1-x)^2}{x}m_\tau^2\bigg).
\eeq
The same form also applies to the HVP contribution, with $\bar \Pi_\ell(s)$ replaced by the corresponding hadronic function. For the numerical results given in the main text, we used the implementation from Refs.~\cite{Keshavarzi:2018mgv,Keshavarzi:2019abf}, and checked that this reproduces the contribution to $a_\tau$ from Ref.~\cite{Keshavarzi:2019abf} quoted in Table~\ref{tab:F2}. Moreover, we checked that Eq.~\eqref{F2lepton} indeed fulfills the dispersion relation
\beq
\label{F2leptondisp}
F_2^\ell(s)=\frac{1}{\pi}\int_{4m_\tau^2}^\infty ds'\,\frac{\Im F_2^\ell(s')}{s'-s},
\eeq
i.e., that the imaginary part generated in the Feynman parameterization via $\Im\bar \Pi_\ell$ reproduces the real part in Eq.~\eqref{F2leptondisp}, where convergence improves when using a subtracted version with $F_2^\ell(0)$ given by Eq.~\eqref{F2lepton0}.

\section{Bremsstrahlung}
\label{sec:bremsstrahlung}

For completeness, we first repeat the expressions~\cite{Barbieri:1972as,Mastrolia:2003yz}
\begin{widetext}
 \begin{align}
  \Re F_1^{(1)}&=-L\bigg(1+\frac{1+y^2}{1-y^2}\log y\bigg)-1-\frac{3y^2-2y+3}{4(1-y^2)}\log y+\frac{1+y^2}{1-y^2}\bigg(\frac{\pi^2}{3}-\frac{1}{4}\log^2 y+\log y\log(1-y)+\Li_2(y)\bigg),\notag\\
  \Im F_1^{(1)}&=-\pi L \frac{1+y^2}{1-y^2}+\pi\bigg[-\frac{3y^2-2y+3}{4(1-y^2)}+\frac{1+y^2}{1-y^2}\bigg(\log(1-y)-\frac{1}{2}\log y\bigg)\bigg],\notag\\
  \Re F_2^{(1)}&=\frac{y}{1-y^2}\log y,\qquad \Im F_2^{(1)}=\pi\frac{y}{1-y^2},\notag\\
  \Re F_2^{(2)}&=-L\bigg[\frac{y}{1-y^2}\log y+\frac{y(1+y^2)}{(1-y^2)^2}\big(\log^2 y-\pi^2\big)\bigg]+(\text{finite}),
 \end{align}
\end{widetext}
where
\beq
y=\frac{\sqrt{s}-\sqrt{s-4m_\tau^2}}{\sqrt{s}+\sqrt{s-4m_\tau^2}},\qquad 
L=\log\frac{\lambda}{m_\tau},
\eeq
and the photon mass $\lambda$ regulates the IR divergence. The finite terms in $\Re F_2^{(2)}$ are lengthy, in the $\lambda$ scheme they can be retrieved from Ref.~\cite{Mastrolia:2003yz}. The IR divergences are canceled by soft emission from the external $\tau$ legs. We calculate these contributions in the soft-photon approximation, since the omitted terms of size $\Order(E_\text{max})$ (with a cutoff $E_\text{max}$ in the photon energy) should be negligible compared to $\sqrt{s}$ and $m_\tau$. The radiative cross section becomes 
\begin{align}
\sigma_\gamma&=\sigma_0\frac{\alpha}{\pi}\eta,\\
\eta&=2\Big(L+\log\frac{m_\tau}{2E_\text{max}}\Big)\bigg(1+\frac{1+y^2}{1-y^2}\log y\bigg)\notag\\
&-\frac{1+y}{1-y}\log y-\frac{1+y^2}{1-y^2}\Big(2\,\Li_2(1-y)+\frac{1}{2}\log^2 y\Big),\notag
\end{align}
which cancels the IR divergence in $2\Re F_1^{(1)}$ in Eq.~\eqref{sig0sig}, as expected. For the remaining IR divergence we collect the singular terms in the numerator of Eq.~\eqref{RF2eff}, which gives 
\begin{align}
\label{ReF2div}
&\Re F_2^{(2)}+\Re F_2^{(1)}\Re F_1^{(1)}+\Im F_2^{(1)}\Im F_1^{(1)}\notag\\
&=-2L\,\Re F_2^{(1)}\bigg(1+\frac{1+y^2}{1-y^2}\log y\bigg)+(\text{finite}).
\end{align}
This divergence is canceled by the same bremsstrahlung diagrams as before, from the terms proportional to $\Re(F_2 F_1^*)=\frac{\alpha}{\pi}\Re F_2^{(1)}+\Order(\alpha^2)$ in the squared matrix element, leading to a correction
\beq
\Re F_2^{\text{eff}, \gamma}=\Big(\frac{\alpha}{\pi}\Big)^2 \Re F_2^{(1)} \eta
\eeq
to be added to Eq.~\eqref{Reff_exp}. This cancels the divergence in Eq.~\eqref{ReF2div}.

In the same way that IR divergences cancel, so does any potential gauge dependence. Defining the $F_2$ form factor itself in a gauge-invariant way for $s>0$ becomes a subtle matter in non-Abelian theories, see, e.g., Refs.~\cite{Papavassiliou:1989zd,
Bernabeu:2002nw}, but for $a_\tau$ these subtleties do not play a role: QCD only enters via HVP, which is manifestly gauge invariant, and the EW contribution is so small that decoupling corrections $\propto s/M_Z^2$ can be neglected. If this were not the case, a potential gauge dependence would have to cancel at the level of $\Re F_2^\text{eff}$, since being defined directly in terms of observables.

\section{Leptoquark model \texorpdfstring{$\boldsymbol{S_1}$}{}}
\label{sec:S1} 

We define the LQ couplings via the Lagrangian
\begin{equation}
 {\mathcal L}=  \left(\lambda_{fi}^{R}\,\bar{u}^c_f\ell_{i}+\lambda_{fi}^{L}\,\bar{Q}_{f}^{\,c}i\tau_{2}L_{i}\right) S_{1}^{\dagger}+\text{h.c.}, 
\end{equation}
where $S_1$ is the scalar LQ $SU(2)_L$ singlet, $f$ and $i$ are flavor indices, $Q$ ($u$) and $L$ ($\ell$) refer to quark and lepton $SU(2)_L$ doublets (singlets), $c$ labels charge conjugation, and $\tau_2$ the second Pauli matrix. As we are only interested in top-quark effects related to $\tau$ leptons, we can set $f=t$, $i=\tau$ and assume the CKM matrix to be diagonal, such that $\lambda_{t\tau}^{R/L}$ are the couplings constrained in Fig.~\ref{fig:atauLQ}.  
Taking into account the leading $m_t/m_\tau$ and $m_t^2/M_Z^2$ enhanced effects, respectively, one finds
\begin{align}
a_\tau^\text{BSM}&=  - \frac{N_c}{48\pi^2}\frac{m_\tau m_t}{M^2}\Re\big[\big(\lambda^R_{t\tau}\big)^*\lambda_{t\tau}^L\big]
\Big(7 + 4\log \frac{m_t^2}{M^2} \Big),\notag\\
\xi_{h\tau\tau}&=
 \bigg| 1 + \frac{m_t}{m_\tau}\frac{N_c}{64\pi ^2}\frac{m_t^2}{M^2}\big(\lambda^R_{t\tau}\big)^*\lambda_{t\tau}^L\notag\\
 &\times\bigg( 2\Big( \frac{M_H^2}{m_t^2} - 4 \Big)\log \frac{m_t^2}{M^2} - 8 + \frac{13}{3}\frac{M_H^2}{m_t^2}\bigg)\bigg|^2,\notag\\
\frac{g_A^\tau}{g_A^\tau\big|_\text{SM}} &= 
1 + \frac{N_c}{16\pi ^2}\Big(\big|\lambda^R_{t\tau}\big|^2+\big|\lambda^L_{t\tau}\big|^2\Big)
\frac{m_t^2}{M^2}\Big(1 + \log \frac{m_t^2}{M^2} \Big),
\end{align}
where $\xi_{h\tau\tau}=\Br\left[h \to \tau ^ +\tau ^ - \right]/\Br\left[ h \to \tau ^ + \tau ^ - \right]|_\text{SM}$, 
$N_c=3$ is the number of colors, $m_t$, $M_H$, $M_Z$, and $M$ are the masses of top quark, Higgs boson, $Z$ boson, and LQ, respectively, and $g_A^\tau$ denotes the axial-vector coupling of the $\tau$~\cite{ParticleDataGroup:2020ssz}.

\bibliography{tau}

\end{document}